\documentstyle[12pt]{article}

\newcommand{\bea}{\begin{eqnarray}}
\newcommand{\eea}{\end{eqnarray}}
\newcommand{\beq}{\begin{equation}}
\newcommand{\eeq}{\end{equation}}

\def\msbar{\ifmmode{\overline{\rm MS}} \else{$\overline{\rm MS}$} \fi}
\def\drbar{\ifmmode{\overline{\rm DR}} \else{$\overline{\rm DR}$} \fi}
\def\drbarr{\ifmmode{\overline{\rm DR}'} \else{$\overline{\rm DR}'$} \fi}

\begin{document}

\hfill\vbox{\baselineskip14pt
            \hbox{LTH-336}
            \hbox{NUB-3094-94TH}
            \hbox{KEK-TH-404}
            \hbox{hep-ph/9407291}
            \hbox{July 1994}}
\vspace{8mm}

\baselineskip22pt
\begin{center}
\Large{Decoupling of the $\epsilon$-scalar mass in softly broken supersymmetry}
\end{center}
\vspace{8mm}

\begin{center}
\large  Ian~Jack, D.~R.~Timothy~Jones,
\end{center}
\begin{center}
{\it DAMTP, University of Liverpool, Liverpool L69 3BX, U.K.}
\end{center}

\vspace{0mm}

\begin{center}
\large  Stephen~P.~Martin, Michael~T.~Vaughn
\end{center}
\begin{center}
{\it Department of Physics, Northeastern University, Boston,
MA 02115, USA}
\end{center}
\begin{center}
and
\end{center}
\begin{center}
\large  Youichi~Yamada\footnote{Fellow of the Japan Society for the
Promotion of Science}
\end{center}
\begin{center}
{\it Theory Group, KEK, Tsukuba, Ibaraki 305, Japan}
\end{center}

\vspace{8mm}
\begin{center}
\large Abstract
\vspace{10pt}
\end{center}
\begin{center}
\begin{minipage}{14cm}
\baselineskip=22pt
\noindent
It has been shown recently that the introduction of an unphysical
$\epsilon$-scalar mass $\tilde{m}$ is necessary for the proper renormalization
of softly broken supersymmetric theories by dimensional reduction ($\drbar$).
In these theories, both the two-loop $\beta$-functions of the scalar masses
and their one-loop finite corrections depend on $\tilde{m}^2$. We find,
however, that the dependence on $\tilde{m}^2$ can be completely removed by
slightly modifying the \drbar renormalization scheme. We also show that
previous \drbar calculations of one-loop corrections in supersymmetry which
ignored the $\tilde{m}^2$ contribution correspond to using this modified
scheme.
\end{minipage}
\end{center}
\vfill
\newpage

\normalsize
\baselineskip=22pt

The usual regularization procedure for the renormalization of softly
broken  supersymmetric theories is dimensional reduction \cite{dr} since
it respects supersymmetry (SUSY), modulo possible high-loop ambiguities
which will not concern us here\cite{sacv}.  In this regularization, an
originally 4-dimensional vector field is  split into two fields: the
first  $D=4-2\epsilon$ components (a $D$-dimensional vector) and the
last  $2\epsilon$ components ($\epsilon$-scalars). All momentum integrals
are $D$-dimensional. The
$\epsilon$-scalars are differently renormalized in non-supersymmetric
theories,  and their interactions, called evanescent couplings, should
in general  be treated as independent of the corresponding interactions
of the vector fields.

In the renormalization of softly broken supersymmetric theories,
all the dimensionless
evanescent couplings are related to the vector couplings by SUSY and
are therefore not independent. For the soft SUSY breaking terms, however,
there is one possible independent
evanescent coupling, the mass $\tilde{m}$ of the
$\epsilon$-scalars.~\footnote{If there is a chiral multiplet which
transforms according to
a representation of the gauge group contained in the
symmetric product of two adjoints,
then there is also a cubic interaction which must be considered\cite{jj}.
 This possibility does not arise in the supersymmetric standard model,
however, so we ignore it here for simplicity.}
Moreover, we cannot set $\tilde{m}=0$
in the renormalization group analysis\cite{jj}, and the two-loop
$\beta$ function of scalar masses in the \drbar scheme
(dimensional reduction with modified minimal subtraction\cite{msbar})
depends explicitly on $\tilde{m}^2$.
In this paper, however, we will show that the $\tilde{m}^2$ dependence
can be removed
from both the two-loop $\beta$ functions of the physical couplings and
the corresponding one-loop finite corrections by a slight modification of
the renormalization scheme. Therefore in this modified scheme
we need not consider the explicit value of $\tilde{m}^2$.

We first present the problem of the $\epsilon$-scalar masses.
Let us consider a supersymmetric theory with the gauge group a product
of simple [or $U(1)$] factors $G_A$, each with a
vector supermultiplet $(V_{\mu}^{Aa}, \chi^{Aa})$ (the index $a$ runs over
the adjoint representation of $G_A$). There are also chiral
supermultiplets $\Phi_i=(\phi_i, \psi_i)$ in representations (reducible,
in general) of $G_A$ with generators $(R^{Aa})_i^j$.
The superpotential $W$ and the soft supersymmetric
breaking terms ${\cal L}_{soft}$ are
\beq
W=\frac{1}{6}\lambda^{ijk}\Phi_i\Phi_j\Phi_k + \frac{1}{2}\mu^{ij}\Phi_i\Phi_j
\eeq
and
\beq
{\cal L}_{soft}=-(m^2)_i^j\phi^{*i}\phi_j-
\left( \frac{1}{6}h^{ijk}\phi_i\phi_j\phi_k+\frac{1}{2}b^{ij}\phi_i\phi_j
+\frac{1}{2}M_A\chi^{Aa}\chi^{Aa}+{\rm h.c.}\right)
\eeq
respectively. (We assume there are no gauge singlet chiral
supermultiplets so there are no terms linear in $\phi$.)
As shown in Ref.\cite{jj}, use of dimensional reduction means that
we have to add an independent evanescent coupling
\beq
{\cal L}_{\epsilon}= \frac{1}{2}\tilde{m}_A^2V^{Aa}_{\sigma}V^{Aa\sigma},
\eeq
the mass term for the $\epsilon$-scalars, to the usual ${\cal L}_{soft}$.
Here the index $\sigma$ runs over the last $2\epsilon$ components
of the original 4-dimensional vector index.

The $\epsilon$-scalar mass term gives contributions to the
next-to-leading order corrections to the scalar masses.
Firstly, the two-loop $\beta$ function for $m^2$
is given by \cite{mv,yy,jj}
\bea
(4\pi)^4\beta^{(2)}_{\drbar}(m^2)_i^j&=&
16\delta_i^j g_A^2g_B^2C_A( i)C_B( i)(2|M_A|^2+M_AM_B^*) \nonumber \\
&&+24\delta_i^j g_A^4C_A(i)[T_A(\Phi)-3C_A(V)] |M_A|^2 \nonumber \\
&&+g_A^2[2C_A(l)-C_A(i)]
[\lambda^*_{ikl}\lambda^{i'kl}(m^2)_{i'}^j
+\lambda^*_{j'kl}\lambda^{jkl}(m^2)_i^{j'}] \nonumber \\
&&-\frac{1}{2}[
\lambda^*_{ikl}\lambda^{lst}\lambda^*_{qst}\lambda^{i'kq}(m^2)_{i'}^j
+\lambda^*_{j'kl}\lambda^{lst}\lambda^*_{qst}\lambda^{jkq}(m^2)_i^{j'} ]
\nonumber \\
&& +2g_A^2[2C_A(l)-C_A(i)][
2\lambda^*_{ikl}\lambda^{jkl}|M_A|^2
-h^*_{ikl}\lambda^{jkl}M_A \nonumber \\
&&-\lambda^*_{ikl}h^{jkl}M_A^* +h^*_{ikl}h^{jkl}
+\lambda^*_{ikl}(m^2)_{k'}^k\lambda^{jk'l}
+\lambda^*_{ikl}(m^2)_{l'}^l\lambda^{jkl'}] \nonumber \\
&&-
h^*_{ikl}h^{lst}\lambda^*_{qst}\lambda^{jkq}
-h^*_{ikl}\lambda^{lst}\lambda^*_{qst}h^{jkq}
-\lambda^*_{ikl}h^{lst}h^*_{qst}\lambda^{jkq} \nonumber \\
&& -\lambda^*_{ikl}\lambda^{lst}h^*_{qst}h^{jkq}
-\lambda^*_{ikl}(m^2)_{k'}^k\lambda^{lst}\lambda^*_{qst}\lambda^{jk'q}
-\lambda^*_{ikl}(m^2)_{l'}^l\lambda^{l'st}\lambda^*_{qst}\lambda^{jkq}
\nonumber \\
&& -\lambda^*_{ikl}\lambda^{lst}\lambda^*_{qst}(m^2)_{q'}^q\lambda^{jkq'}
-2\lambda^*_{ikl}\lambda^{lst}(m^2)_t^{t'}\lambda^*_{qst'}\lambda^{jkq}
\nonumber\\
&& -2g_A^2 (R^{Aa})^j_i(R^{Aa} m^2)^l_r\lambda^*_{lpq}\lambda^{rpq}
 +8g_A^2 g_B^2 (R^{Aa})^j_i {\rm Tr}[R^{Aa} C_B(r) m^2]
\nonumber\\
&&+16\delta_i^jg_A^4C_A(i)\left( {\rm Tr}[ C_A(r)m^2]/d(G_A)-C_A(V)|M_A|^2
\right) \nonumber\\
&&+8\delta_i^j g_A^4C_A(i) [T_A(\Phi)- 3C_A(V)]\tilde{m}_A^2
\nonumber\\
&&-2g_A^2[C_A(i)+2C_A(l)]\lambda^*_{ikl}\lambda^{jkl}\tilde{m}_A^2.
\eea
Here the $C_A(i)$ are the
eigenvalues of the Casimir operator defined by
\beq
C_A(i)\delta_i^j = (R^{Aa}R^{Aa})_i^j .
\eeq
$C_A(V)$ is the eigenvalue on the adjoint representation of $G_A$, and
$C_A(r)$ which appears in the trace terms denotes the eigenvalue of $C_A$ on
the irreducible representations $r$ appearing in the trace.
$T_A(\Phi)$ is the Dynkin index
\beq
T_A(\Phi)\delta^{ab} = {\rm Tr}(R^{Aa}R^{Ab})
\eeq
and $d(G_A)$ is the dimension of $G_A$.

In fact the proper inclusion of the $\epsilon$-scalar mass and
its counterterms resolves the discrepancy between the result for
$\beta^{(2)}(m^2)$ in \cite{jj} and those in the original versions
of \cite{mv,yy}.

Secondly, the one-loop pole masses of scalars also depend on
$\tilde{m}^2$. The general expression is:
\beq
m_i^2({\rm pole})=m_i^2|_{\drbar}-\frac{2g_A^2C_A(i)}{(4\pi)^2}\tilde{m}_A^2
-\Pi_i(q^2=m_i^2({\rm pole}), m_i^2).
\eeq
Here $\Pi_i(q^2 , m_i^2)$ is the one-loop two-point function of
$\phi_i$ calculated in \drbar with $\tilde{m}=0$.
Assuming there are no gauge singlets,
there are no other physical $\beta$
functions or finite corrections which depend on $\tilde{m}^2$.

Since the $\epsilon$-scalar mass is not an observable, the results
(4) and (7) appear at first sight somewhat paradoxical. This issue was
addressed in Ref.\cite{jjk}, where it was proved that \drbar is related to
the standard dimensional regularisation scheme ($\msbar$)
 \cite{dreg} by coupling
constant redefinition. It follows that the S-matrix is identical in the two
schemes. Nevertheless, the $\tilde{m}^2$ dependence is
an inconvenience for the \drbar scheme.
Although it is safe simply to impose $\tilde{m}=0$ for calculations at a
fixed renormalization scale, this condition is unstable under
the renormalization group and may not be used in the renormalization
group analyses. This is obvious from the one-loop $\beta$ function of
$\tilde{m}^2$ \cite{jj},
\beq
(4\pi)^2\beta^{(1)}_{\drbar}(\tilde{m}_A^2)=
2g^2_A[T_A(\Phi)-3C_A(V)]\tilde{m}_A^2
+4g_A^2({\rm Tr}[C_A(r) m^2]/d(G_A)-C_A(V)|M_A|^2),
\eeq
which is inhomogeneous with respect to $\tilde{m}^2$.

One could of course use the \msbar scheme, which is manifestly
$\tilde{m}^2$--independent,  but \msbar does not respect SUSY and so a
much more complicated treatment of the physical couplings would be
entailed. What we would like is the best of both worlds: a prescription
which respects SUSY  and yet contains no $\tilde{m}^2$ dependence. One
might expect that there is such a prescription,  since  $\tilde{m}^2$ is
an unphysical artifact produced by dimensional reduction.  We find that
this is indeed the case.  The $\tilde{m}^2$ dependences of (4) and (7)
are removed by a simple modification of the renormalization scheme from the
\drbar scheme. Let us consider a new scheme ($\drbarr$) defined as
\beq
(m^2)_i^j|_{\drbarr}=(m^2)_i^j|_{\drbar}-\frac{2
g_A^2C_A(i)}{(4\pi)^2}\delta_i^j
\tilde{m}_A^2
\eeq
while all other couplings are not modified from $\drbar$.  In fact,
$m^2|_{\drbarr}$ is the same  as the  corresponding object in the
dimensional  regularization scheme ($\msbar$). The reason for this
becomes  clear with  the realisation that we could arrive at $\drbarr$
starting from $\msbar$, by  making the redefinitions necessary to effect
the change from $\msbar$ to  $\drbar$
for all couplings {\it except} the scalar mass $(m^2)_i^j$.
These redefinitions were given in \cite{mvo}.  Since $\msbar$ is manifestly
$\tilde{m}^2$--independent, it follows that this scheme ($\drbarr$) will
be too.

In the \drbarr scheme, the two-loop $\beta$ function for $m^2$ becomes
\bea
(4\pi)^4\beta_{\drbarr}^{(2)}(m^2)_i^j&=
&(4\pi)^4\beta_{\drbar}^{(2)}(m^2)_i^j)\nonumber\\
&&-8\delta_i^jg_A^4C_A(i)\left( {\rm Tr}[C_A(r)m^2]/d(G_A)-C_A(V)|M_A|^2
\right) \nonumber\\
&&-8\delta_i^jg_A^4C_A(i) [T_A(\Phi)-3C_A(V)]\tilde{m}_A^2
\nonumber\\
&&+2g_A^2[C_A(i)+2C_A(l)]\lambda^*_{ikl}\lambda^{jkl}\tilde{m}_A^2.
\eea
This $\beta^{(2)}_{\drbarr}(m^2)$, which has no $\tilde{m}^2$ dependence,
agrees with the result of Ref.\cite{mv}. Eq.~(10) follows from (8), (9),
and the one-loop $\beta$ function\cite{1loopm}
\bea
(4\pi)^2\beta^{(1)}(m^2)_i^j&=&
\frac{1}{2}[\lambda^*_{ikl}\lambda^{i'kl}(m^2)_{i'}^j
+\lambda^*_{j'kl}\lambda^{jkl}(m^2)_i^{j'}]
+2\lambda^*_{ikl}(m^2)_{l'}^l\lambda^{jkl'}  \nonumber \\
&&
+h^*_{ikl}h^{jkl} -8g_A^2C_A( i)|M_A|^2\delta_i^j
+2g_A^2(R^{Aa})_i^j{\rm Tr}(R^{Aa}m^2).
\eea
The one-loop pole masses of scalars, given by
\beq
m_i^2({\rm pole})=m_i^2|_{\drbarr}-\Pi_i(q^2=m_i^2({\rm pole}), m_i^2),
\eeq
are also $\tilde{m}^2$ independent.

The sets of $(m^2, \tilde{m}^2)_{\drbar}$
which give the same $m^2|_{\drbarr}$ are of course physically equivalent.
By comparing (12) with (7), we can also see that
all the previous calculations of
one-loop mass corrections in SUSY,
which have ignored the contribution of
$\epsilon$-scalar mass, are then
justified as calculations in this \drbarr scheme.

In summary, we have found that the $\epsilon$-scalar mass dependence
of the two-loop $\beta$ functions and of one-loop finite
corrections can be completely removed by a slight modification of the
renormalization scheme from the \drbar scheme.
We have also shown that the previous calculations of the one-loop
mass corrections which have ignored the $\epsilon$-scalar mass
contribution are justified as being (unwittingly) calculations
in this new renormalization scheme, $\drbarr$.

\section*{\large Acknowledgements}

Y.~Y. would like to thank K. Hagiwara for fruitful discussions.
The work of D.~I.~J.~ and D.~R.~T.~J.~is supported in part by PPARC.
The work of S.~P.~M.~and M.~T.~V.~is supported in part by the
National Science Foundation grants PHY-90-01439 and PHY-93-06906 and
U.~S.~Department of Energy grant DE-FG02-85ER40233.
The work of Y.~Y.~is supported in part by the Japan Society for the
Promotion of Science and the Grant-in-Aid for Scientific Research
from the Ministry of Education, Science and Culture
of Japan No.~06-1923.


\def\PL #1 #2 #3 {Phys.~Lett. {\bf#1}, #2 (#3) }
\def\NP #1 #2 #3 {Nucl.~Phys. {\bf#1}, #2 (#3) }
\def\ZP #1 #2 #3 {Z.~Phys. {\bf#1}, #2 (#3) }
\def\PR #1 #2 #3 {Phys.~Rev. {\bf#1}, #2 (#3) }
\def\PP #1 #2 #3 {Phys.~Rep. {\bf#1}, #2 (#3) }
\def\PRL #1 #2 #3 {Phys.~Rev.~Lett. {\bf#1}, #2 (#3) }
\def\PTP #1 #2 #3 {Prog.~Theor.~Phys. {\bf#1}, #2 (#3) }
\def\ib #1 #2 #3 {{\it ibid.} {\bf#1}, #2 (#3) }
\def\etal {{\it et al}.}
\def\eg {{\it e.g}.}
\def\ie {{\it i.e}.}



\begin{thebibliography}{99}

\bibitem{dr}
W.~Siegel, \PL 84B 193 1979 ;\\
D.~M.~Capper, D.~R.~T.~Jones and P.~van Nieuwenhuizen, \NP B167 479 1980 .

\bibitem{sacv}
W.~Siegel, \PL 94B 37 1980 ;\\
L.~V.~Avdeev, G.~A.~Chochia and A.~A.~Vladimirov, \PL 105B 272 1981 .

\bibitem{jj}
I.~Jack and D.~R.~T.~Jones, Liverpool preprint LTH 334 (1994)
(to be published in Phys.~Lett.~B).

\bibitem{msbar}
W.~A.~Bardeen, A.~J.~Buras, D.~W.~Duke and T.~Muta,
\PR D18 3998 1978 .

\bibitem{mv}
S.~P.~Martin and M.~T.~Vaughn, Northeastern preprint
NUB-3081-93TH (1993) (to be published in Phys.~Rev.~D).

\bibitem{yy}
Y.~Yamada, KEK preprint KEK-TH-383 (1994)
(to be published in Phys. Rev. D).

\bibitem{jjk}
I.~Jack, D.~R.~T.~Jones and K.~L.~Roberts, Liverpool preprint LTH 329 (1994)
(to be published in Z.~Phys.~C).

\bibitem{dreg}
G.~'t Hooft and M.~Veltman, \NP B44 189 1972 .

\bibitem{mvo}
S.~P.~Martin and M.~T.~Vaughn, \PL 318B 331 1993 .

\bibitem{1loopm}
K.~Inoue, A.~Kakuto, H.~Komatsu and S.~Takeshita, \PTP 68 927 1982 ;
\ib 70 330 1983(E) ; \ib 71 413 1984 .

\end{thebibliography}
\end{document}